\documentclass[useAMS,usenatbib]{mn2e}

\usepackage{amsmath,amssymb,graphicx}

\newcommand{\kepler}{\textit{Kepler}}
\newcommand{\nhots}{144}

\newcommand{\nmult}{112}
\newcommand{\nnew}{24}
\newcommand{\newrate}{17}

\begin{document}


\title{A population of planetary systems characterized by short-period, Earth-sized planets}

\author[Steffen \& Coughlin]{
Jason H. Steffen$^1$ and Jeffrey L. Coughlin$^{2,3}$ \\
$^1$University of Nevada, Las Vegas, Department of Physics and Astronomy, 4505 S Maryland Pkwy, Box 454002, Las Vegas, NV 89154-4002 \\
$^2$SETI Institute, 189 Bernardo Ave, Suite 200, Mountain View, CA 94043, USA \\
$^3$NASA Ames Research Center, M/S 244-30, Moffett Field, CA 94035, USA
}


\pagerange{\pageref{firstpage}--\pageref{lastpage}} 

\maketitle

\label{firstpage}

\begin{abstract}
We analyze data from the Quarter 1-17 Data Release 24 (Q1--Q17~DR24) planet candidate catalog from NASA's \kepler\ mission, specifically comparing systems with single transiting planets to systems with multiple transiting planets, and identify a distinct population of exoplanets with a necessarily distinct system architecture.  Such an architecture likely indicates a different branch in their evolutionary past relative to the typical \kepler\ system.  The key feature of these planetary systems is an isolated, Earth-sized planet with a roughly one-day orbital period.  We estimate that at least \nnew\ of the \nhots\ systems we examined ($\gtrsim$\newrate\%) are members of this population.  Accounting for detection efficiency, such planetary systems occur with a frequency similar to the hot Jupiters.
\end{abstract}

\bigskip

\section*{Introduction}

Modern surveys of exoplanets have produced a variety of planetary systems with properties that are often quite distinct from the Solar System\citep{Winn:2015}.  The architectures of these systems, the orbits and masses of the planets and how those quantities relate to each other, indicate potential differences in their formation and evolutionary histories \citep{Lissauer:2011b,Rasio:1996,Lee:2002,Fabrycky:2007,Dawson:2015}.  These histories, in turn, give insights into the various processes at work in the formation and subsequent dynamical evolution of the system.  Identifying different planet populations and the occurrence rates of those populations are, therefore, key ingredients to generalizing our planet formation models.  Hot Jupiters, for example, are a distinct population of planets with histories that differ from either the Solar System or most of the planetary systems discovered to date \citep{Gaudi:2005,Cumming:2008,Wright:2009,Steffen:2012b}.

Several studies of the architectures of planetary systems discovered by NASA's \kepler\ mission have noted differences between the single-planet systems and the multi-planet systems \citep{Latham:2011,Lissauer:2011b,Hansen:2013,Moriarty:2015}.  For example, multiplanet systems rarely contain hot Jupiters with nearby planetary companions \citep{Wright:2009,Steffen:2012b}---thus far, WASP-47 is the only exception to this rule \citep{Becker:2015}.  Studies comparing the relative frequencies of systems with two, three, and higher multiplicity systems show that planetary systems typically have multiple planets whose orbits are mutually inclined by a few degrees \citep{Lissauer:2011b,Fang:2012,Hansen:2013}.  However, these models also indicate an excess of observed single-planet systems when mutual inclinations are modeled by simple distributions \citep[e.g., a Rayleigh distribution,]{Lissauer:2011b,Fang:2012,Hansen:2013,Moriarty:2015}.  While this discrepancy can be mitigated to an extent with more sophisticated, single parameter models \citep{Lissauer:2011b}, additional investigations comparing single planet and multiplanet systems are warranted as a means to identify and characterize the source of the possible excess of single planet systems and to uncover populations of planets with unique system architectures.

Here we investigate differences between the properties of the singles and multis given in the most recent \kepler\ planet candidate catalog in an effort to identify new planet populations that may be present among the singles that do not appear in similar numbers among the multis.  Should such populations exist, they could contribute to the putative excess of single-planet systems.  The distinguishing characteristics of these systems may give insights into their origins and, hence, into when and how the evolutionary path of the new population diverged from that which produced the more typical Kepler-like systems.

This work will proceed as follows.  We first describe our sample selection process and outline a Monte Carlo study of the singles and multis to identify regions of interest in planet parameters.  For one of these regions, we estimate the contribution two possible sources for the observations---namely false positives and non-transiting planets.  We conclude by discussing possible origins and implications of these results.

\section*{Comparison of Singles and Multis}\label{montecarlo}

We select our sample of planet candidates from the Q1--Q17~DR24 \kepler\ planet candidate catalog \citep{Coughlin:2015b}.  This is the first such catalog to have robotically and uniformly evaluated all planetary signals detected by the \kepler\ pipeline.  In this catalog, a wide range of false positives are identified, including instrumental artifacts, spotty stars, pulsating stars, eclipsing binaries, and contamination from variable stars due to crowding, stray light, and detector effects.  This false positive identification is based solely on \kepler\ data, i.e., no ground-based follow-up observations are used.  We construct our sample starting with the 4,293 planet candidates, plus 9 confirmed planets that were misclassfied as false positives \citep[][\S5.2.5]{Coughlin:2015b}, in the Q1--Q17~DR24 catalog.  The completeness of this catalog is investigated in detail in \citep{Coughlin:2015b,Christiansen:2015}.

\begin{figure}
\centering
\includegraphics[width=.9\linewidth]{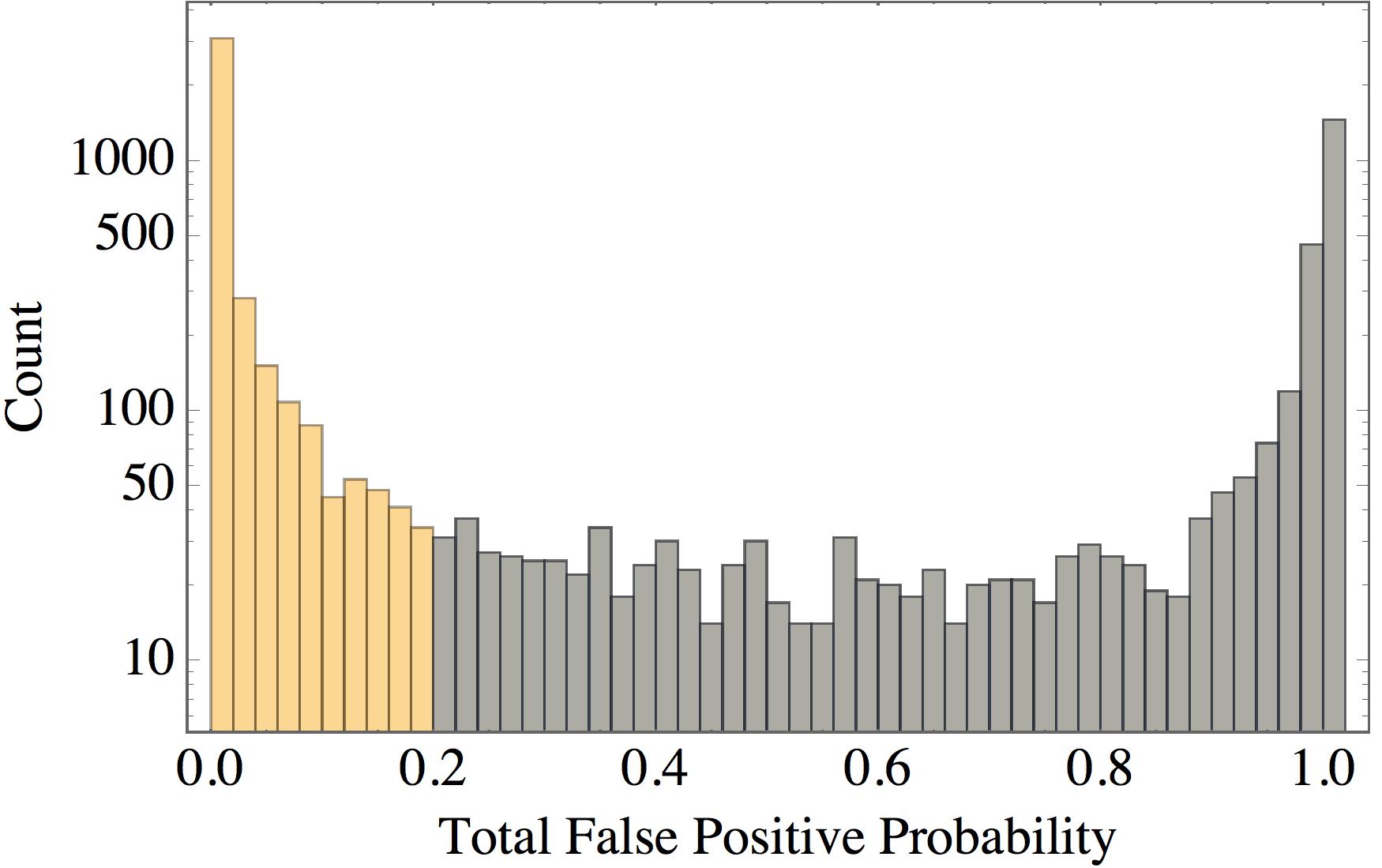}
\caption{Total astrophysical false positive probabilities for all KOIs as determined by the VESPA software.  Our sample of systems was restricted to those where the maximum, total false positive probability for any candidate in the system was greater than 0.2.}
\label{fpplot}
\end{figure}

Beyond the vetting done in producing the catalog, each candidate has been analyzed using the VESPA code \citep{Morton:2012} to determine the probability that the transit-like signal is actually a false positive caused by an undetected eclipsing binary star.  These scenarios can include such cases as an eclipsing binary that has a secondary eclipse depth below the noise level, or one that has nearly identical eclipse depths and is detected at half the orbital period.  The distribution of the total probability that the planet candidates are among the considered false positive scenarios are shown in Figure \ref{fpplot}.  We obtain the VESPA false positives probabilities from the NASA Exoplanet Archive, and remove any candidate system (including singles) that was either not vetted using VESPA or where the total false positive probability is greater than 0.2 for the worst-case candidate within the system.  Our resulting sample includes 3063 candidates in 2373 systems---1890 in single systems, and 1173 candidates in 483 multiplanet systems.

\begin{figure}
\centering
\includegraphics[width=.9\linewidth]{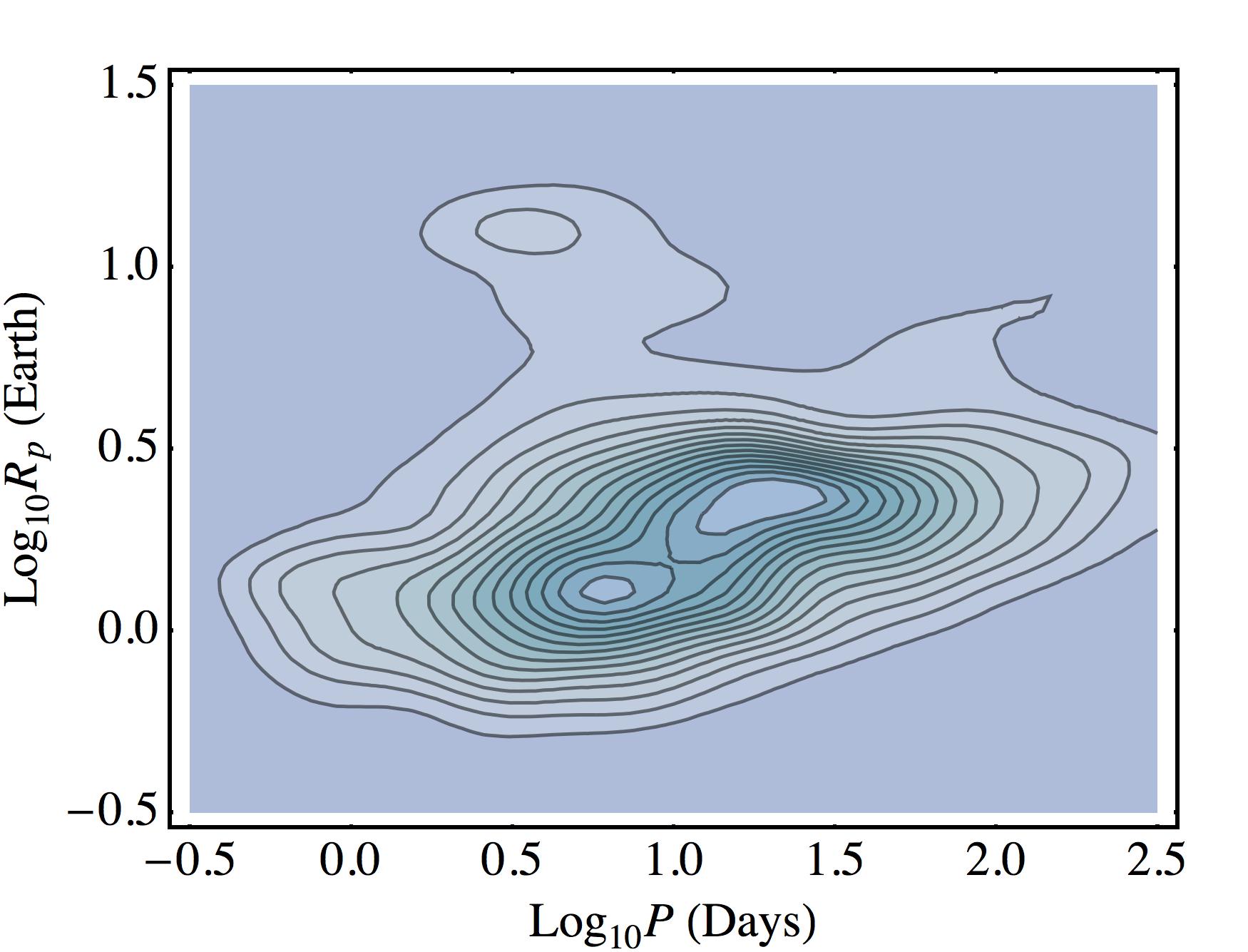}
\includegraphics[width=.9\linewidth]{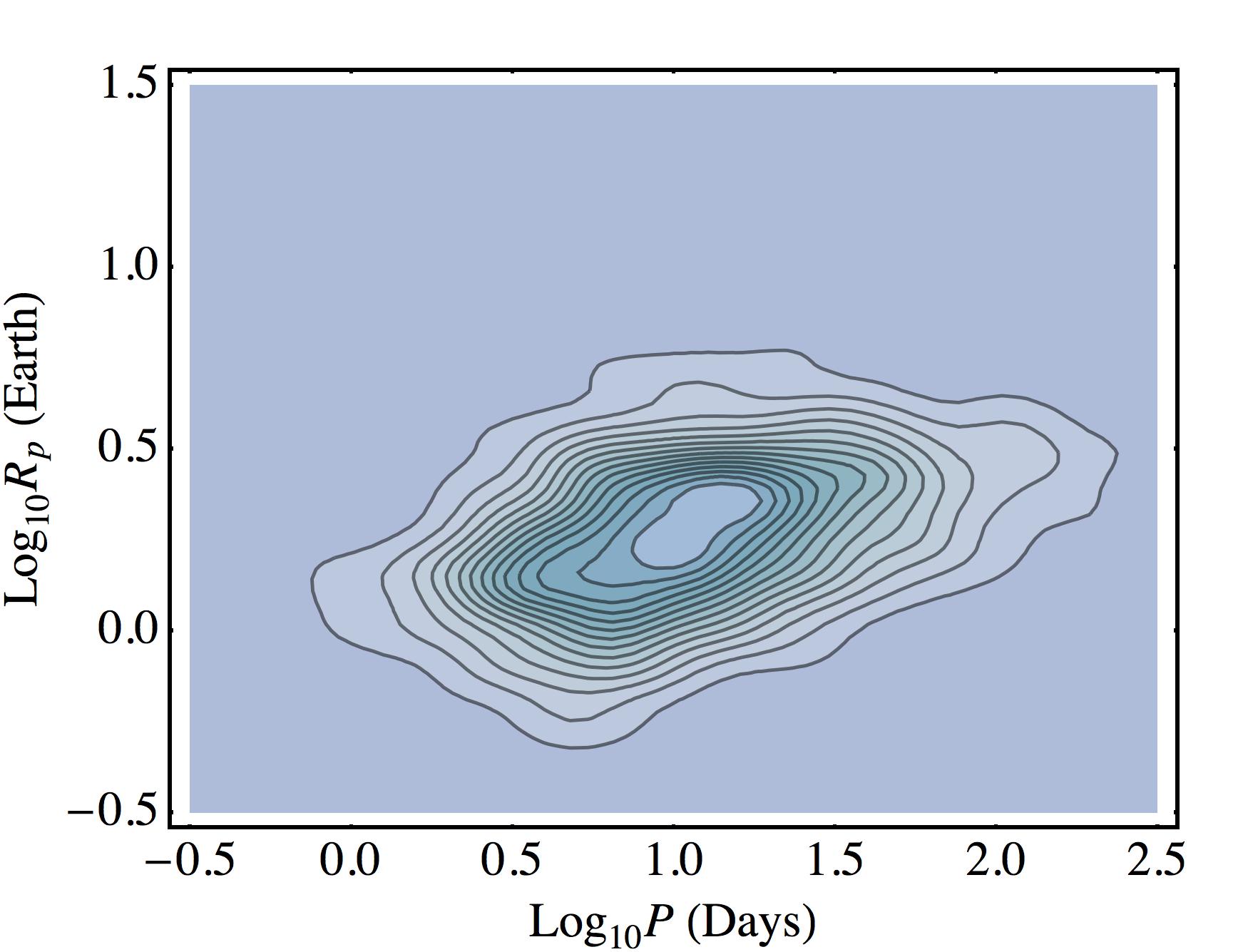}
\caption{Kernel density estimates for the distribution of observed planet candidate sizes and orbital periods for single planet systems (top) and multiplanet systems (bottom).  The area under each distribution is normalized to unity.}
\label{kdes}
\end{figure}

The panels in Figure \ref{kdes} show a kernel density estimation (KDE) of the distribution of planet size and orbital period for the single and multiplanet systems.  If the observed single planet systems are actually multiplanet systems where only one planet happens to transit, and if we assume that the detection efficiency is similar for both groups, then we should be able to reconstruct the size/period distribution of singles by randomly drawing samples from the multis.  Regions where the sample multis are unable to consistently reproduce the observed frequency of singles merit further study as they may indicate the presence of distinct planet populations.

\begin{figure}
\centering
\includegraphics[width=.9\linewidth]{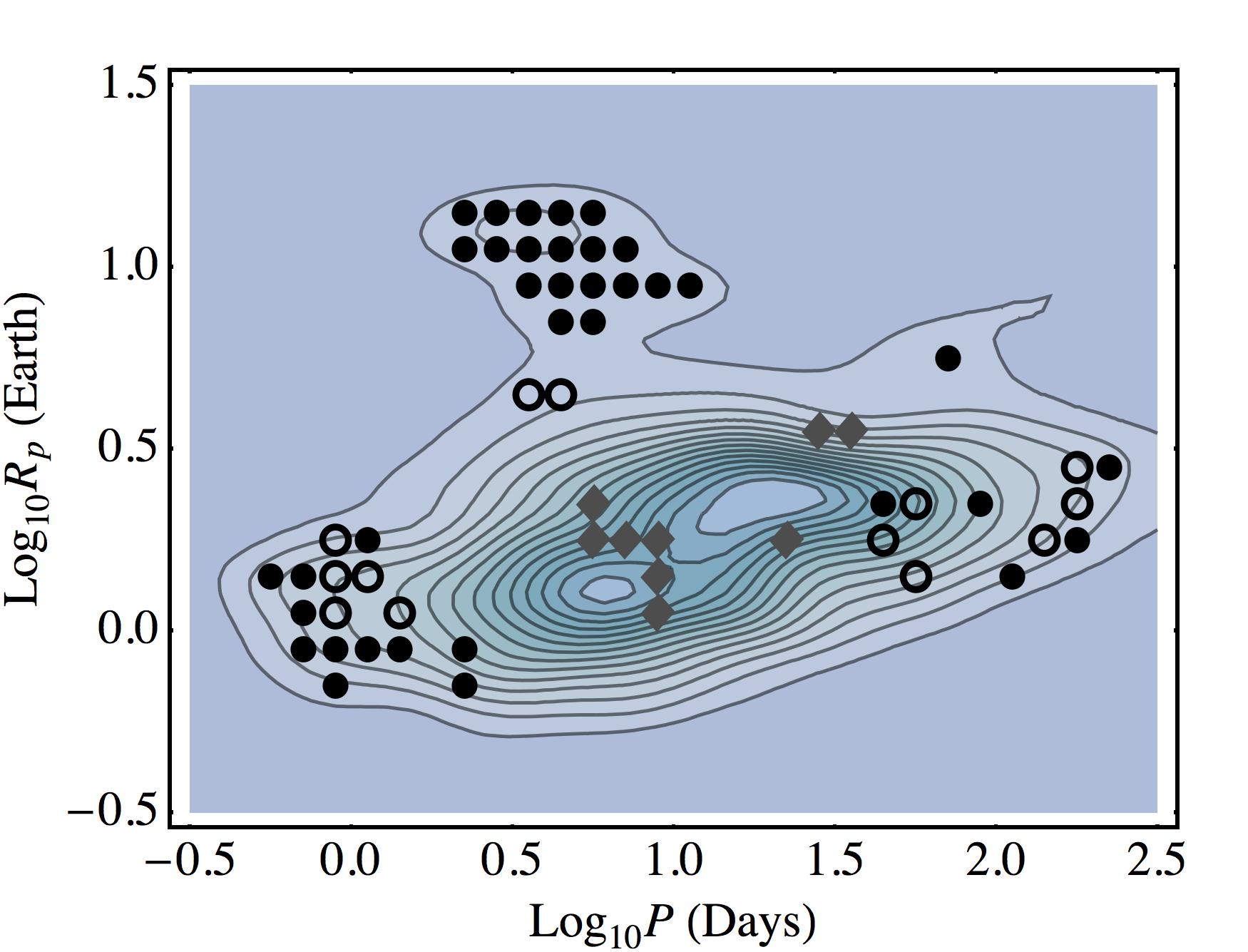}
\caption{Results of a Monte Carlo simulation to reproduce the single planet systems using a random sampling of 1890 realizations drawn from the multiplanet systems.  Black disks and circles correspond to bins where the multiplanet systems systematically underproduce the observed number of single planet systems (less than 2.5\% and 0.5\% of the realizations respectively).  Grey diamonds correspond to bins where the multiplanet distribution systematically overproduces the observed number of singles 97.5\% of the time).  These regions where the multiplanet systems do not match the properties of the single planet systems indicate potentially new populations of planetary system.  Our study focuses primarily on those near one Earth radius and one-day orbital periods.}
\label{mcresults}
\end{figure}

We ran 100,000 realizations of a Monte Carlo simulation where 1890 random draws from the KDE for multis are compared with the observed singles in logarithmically uniform bins in planet size and orbital period.  The results of this simulation are shown in Figure \ref{mcresults}.  Bins with black circles and disks indicate where less than 2.5\% and 0.5\% of the realizations reproduce the singles respectively.  Gray diamonds are where the multis overproduce singles 97.5\% of the time.

There are four regions of interest where the marked bins cluster.  For three of the regions, the distribution of multis consistently underproduces the number of observed singles.  For one region, the multis consistently overproduce them.  The region overproduced by the multiplanet distribution is for planets a few times the size of the Earth with orbits of tens of days.  The underproducing regions include the hot Jupiters with radii near 10 R$_\oplus$ and orbits near three days, planets with radii of 2--4 R$_\oplus$ on wide, $\sim 100$-day orbits, and Earth-sized planets on one-day orbits (hereafter ``hot Earths'').

The first three regions we discuss briefly, but do not investigate in detail.  First, hot Jupiters are a known population with a unique system architecture where the planets are either largely isolated or are single \citep{Wright:2009,Steffen:2012b}, thus we would not expect the distribution of multis to reproduce them.  Second, the excess singles at long orbital periods may be explained as the inner planets of larger systems where the outer planets are missed by the limited timespan of the \kepler\ observations.  Or, these candidates may be false positives as there are a relatively large number of spurious signals at long orbital periods \citep{Mullally:2015,Coughlin:2015b}.  Third, the excess of multis near the center of the distribution is likely an artifact of the comparison.  Specifically, since the distribution of singles has larger tails (e.g., it has nonzero probability density in the region of the hot Jupiters), the distribution of multis must have a higher peak in order to be properly normalized.  This peak then systematically overproduces planets in its vicinity, as is seen in Figure \ref{mcresults}.  For the remainder of this work, we will focus on the fourth region of interest, the population of hot Earths on one-day orbits.

\section*{Source budget for hot Earths}\label{budget}

There are four possible sources of hot-Earth candidates.  They may be: 1) statistical false positives---photometric fluctuations that mimic a transit signal, 2) astrophysical false positives such as background eclipsing binary stars, 3) the innermost planets of standard \kepler\ multiplanet systems where the rest of the system is not detected, or 4) they may be planets from systems with a new and distinct architecture---such as single planets or planets that are widely separated from their neighbors.  We calculate the budget for each of the first three sources to determine how many, if any, hot Earths may be members of a new population.  The results of our study, which are discussed in more detail in the following sections, are broken down in Table \ref{budgettable}.

\begin{table}
\centering
\caption{Expected rate and conservative limit on the frequency and number of contributions to the hot Earth population.\label{budgettable}}
\begin{tabular}{lrrr}
\multicolumn{1}{c}{Source} & \multicolumn{1}{c}{Expected} & \multicolumn{1}{c}{Limit} \\
& \multicolumn{1}{c}{rate (number)} & \multicolumn{1}{c}{rate (number)} \\
\hline
Kepler False Positives		& 0.014 (2.0)		& 0.033 $(<5)$  \\
Binary False Positives		& 0.020 (2.8)		& 0.021	$(<3)$  \\
Multiplanet		& 0.54 (77.3)		& 0.78 $(<112)$  \\
\hline
This Population	& 0.43 (61.9)		& 0.17 $(>24)$  \\
\hline
\end{tabular}
{\small Expected rates are from our analysis means, our primary claims use the stated 84\% limit.  For multiplanet systems, this rate itself results from conservative assumptions on inclinations and period ratios.}
\end{table}

\subsection*{Stellar and Statistical False Positives from Kepler Data}

\begin{figure}
\centering
\includegraphics[width=.9\linewidth]{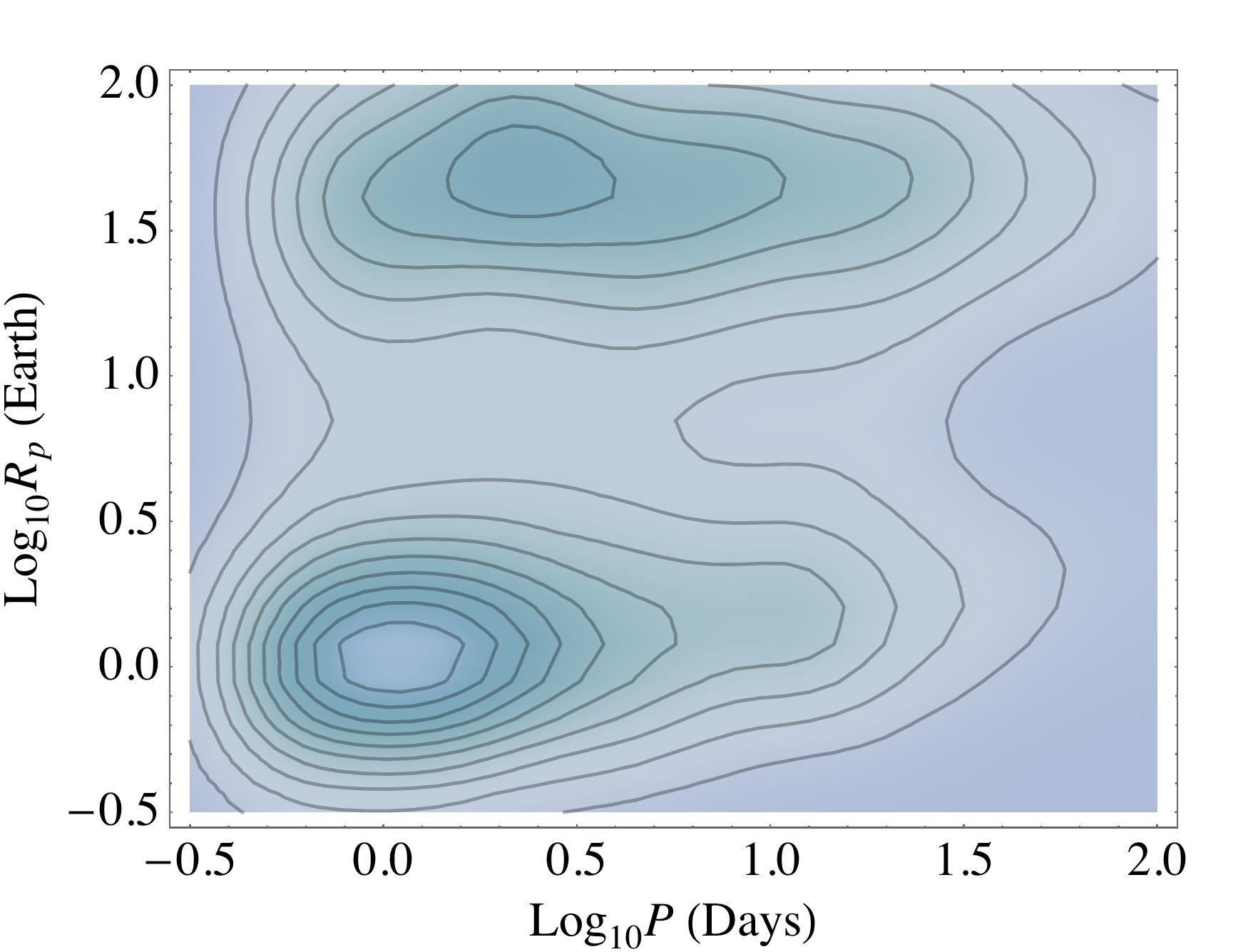}
\caption{Smooth histogram of the distribution of known false positives from the \kepler\ catalog in terms of the logarithm of the inferred planet radius and orbital period.  The lower peak of this distribution is centered at one Earth radius and one-day orbital periods, coincident with our primary region of interest.}
\label{picfalsepos}
\end{figure}

In the planet candidate catalog there are \nhots\ hot-Earth planet candidates with sizes less than 2 R$_\oplus$ and orbital periods less than 2 days. In the same regime, there are 562 known false positives, due both to instrumental artifacts and to astrophysical sources.  Figure \ref{picfalsepos} shows a smooth histogram of the relative number of Kepler Objects of Interest (KOIs) that are dispositioned as false positives in the catalog.  The fact that there is a peak in the false positive distribution in our region of interest motivates extra scrutiny to estimate the rate that they contaminate our candidate sample.

We use the Threshold Crossing Event Review Team vetting forms \citep{Coughlin:2015a}, which are publicly available at the NASA Exoplanet Archive, to manually examine the \nhots\ planet candidates.  From these reports we find that only two are likely false positives: KOI 6648.01 is due to a variable star, and 7051.01's signal originates from a nearby bright star.  The rest of the \nhots\ candidates show no evidence from \kepler\ data alone that they are due to variable stars, EBs, systematics, or any other known false positive source.  We note that five of them (KOIs 2916.01, 4144.01, 4441.01, 4685.01, and 6262.01), while they are viable planet candidates, have orbital periods that are half of the period listed DR24 catalog.  This discrepancy is due to the actual orbital period being less than 0.5 days---the lower search limit of the \kepler\ pipeline.

Additionally, we examined the Kepler False Positive Table \citep{Bryson:2015}, which uses data aside from the \kepler\ light curves, such as ground-based Doppler observations, to determine false positive status.  None of the 144 were listed as false positives in the table.  Also, as part of the currently on-going DR25 analysis, a \kepler\ pipeline search was conducted after inverting the input light curves.  Candidate detections in the inverted light curve would weaken the candidate's validity as it would likely be due to quasi-sinusoidal brightness variations.  None of the \nhots\ were detected as planet candidates in this preliminary DR25 inverted pipeline search.  Based on these results, we estimate the false alarm contamination rate for our hot-Earth sample as 1.4\% (2/144).  Assuming Poisson statistics, this yields an 84\% upper limit on the rate of 3.2\% ($<5/144$).


\subsection*{Eclipsing Binary False Positives}

We estimate the number of false positives from eclipsing binaries that may yet remain among the planet candidates using the VESPA analysis of the candidate systems \citep{Morton:2012}.  VESPA identifies the probability that a transit signal is due to one of six possible sources: hierarchical eclipsing binaries, blended eclipsing binaries (eclipsing binaries where an unassociated third star dilutes the eclipse signal), unblended eclipsing binaries, and these same three scenarios but with the binary period at twice the planet candidate period.  (The primary and secondary eclipses of a binary system, if they have similar depth, can be misinterpreted as a planet candidate with half the binary orbital period.)

Summing the probability of each false positive scenario over all \nhots\ candidates in our sample yields an estimate of 2.8 false positives (1.9\%).  The fractional uncertainty in this quantity is $1/\sqrt{\nhots} \simeq 8.3$\%, or $\pm 0.2$ planets.  Using an 84\% upper limit on the false positive rate, we expect less than 2\% (or 3 planets) of the sample to be false positives of these types.

\subsection*{Standard Multiplanet Systems}

Some fraction of the hot Earths are likely members of multiplanet systems similar to the bulk of the systems in the \kepler\ catalog---having nearby planetary companions with period ratios $\lesssim 3$.  (The fact that the hot Earths are systematically smaller than the typical \kepler\ planet may simply be due to photoevaporation of the atmosphere \citep{Baraffe:2004,Valencia:2010,Lopez:2013a,Owen:2013}.)  We estimate the number of hot Earths that may be members of such systems using the method employed by Steffen and Hwang \citep{Steffen:2015} to calculate the number of systems that where only a single transiting hot Earth would be seen for each hot Earth that is observed in a multiplanet system.  This approach corrects for both the geometric effect where outer planets do not transit the host star and pipeline completeness where more distant planets are harder to detect.  It uses the distance from the inner planet to the star, the mutual inclination of the orbits, and the ratio of the planetary orbital periods as given in the \kepler\ data products \citep{Coughlin:2015b}.

For this analysis we estimate that the mutual inclinations are Rayleigh distributed with a Rayleigh parameter of 1.5 degrees---a value that is on the low side of existing estimates \citep{Lissauer:2011b,Fang:2012,Tremaine:2012}.  This choice of low mutual inclination orbits is conservative as follows.  When mutual inclinations are small, then planets are on wide orbits will have an increased likelihood that an outer planet will transit, given that an inner planet transits.  However, this effect changes once the planet orbital distances cross the threshold where the mutual inclinations are larger than the ratio of the stellar size to the orbital distance ($R_\star / a \ll i$).  In this regime it is less likely that the outer planet will transit because grazing transits of the inner planet would yield outer planets whose orbits are projected well above or below the stellar disk.  On the other hand, with these short orbital periods, larger mutual inclinations enable more distant planets to transit.  In our budget, having low mutual inclinations will (conservatively) overestimate the number of predicted outer planets given the observations.


For our analysis we use observed values for the ratio $R_\star / a$.  We also assume orbital period ratios equal to those of the adjacent pairs in the observed multiplanet systems but restricted by two criteria.  First, the inner planet of the pair must have an orbital period greater than 5 days so that our results are not biased by potential differences in architecture in the regime we are considering.  There is a known observational difference in observed period ratios with orbital periods less than a few days \citep{Steffen:2013c}, which this criterion avoids.

The second restriction is that the period ratio must be less than 6.4.  This criterion is somewhat arbitrary (it is the largest ratio for adjacent pairs in the Solar System) but our results are not particularly sensitive to this choice.  Moreover, we expect that there will be many intermediate, nontransiting planets in the observed systems \citep{Lissauer:2011b,Fang:2012,Hansen:2013,Tremaine:2012,Steffen:2013b} and thus the observed adjacent period ratios are likely to be larger than the true adjacent period ratios.  Thus, the cut at a ratio of 6.4 prevents the exacerbation of an existing bias our results toward erroneously large period ratios.

We ran a Monte Carlo simulation with 100,000 realizations of planetary period ratio and $R_\star / a$---calculating the number of missed multiplanet systems for each observed hot Earth in a multi (there are 44).  We assume the outer planet to be 12\% larger than the observed hot Earth \citep{Ciardi:2013}---a conservative choice given that the size of the hot Earth is likely smaller than its initial size due to atmospheric evaporation \citep{Valencia:2010,Lopez:2013a,Owen:2013}.  The median number of missed second planets is 1.8 per system (with a median period ratio near 2.0) while the 84\% upper limit on the number of missed planets per system is 2.6 (corresponding to a period ratio near 3.1).  These results yield an estimate of 77 hot Earths from multiplanet systems that are observed as singles and an upper limit of \nmult\ such systems.

\subsection*{Isolated Hot Earths}

From these analyses, summarized in Table \ref{budgettable}, we estimate that no fewer than \nnew\ (about one in six or $\sim$17\%) of the \nhots\ single hot Earth candidates are members of a new population of planetary systems with architectures that are distinctly different from the rest of the \kepler\ sample of planets. These planets may either be truly single, or effectively single planets that are the innermost members of multiplanet systems that have become dynamically isolated from their planetary siblings (e.g., having period ratios $\gtrsim 10$).

One possible source of these hot Earths is left-over cores of hot Jupiters following the loss of their atmospheres through Roche lobe overflow \citep{Valsecchi:2014}.  Another example of the kind of system that may contribute to the hot Earth sample is the Kepler-10 system \citep{Batalha:2011}.  This system comprises a 1.5 R$_\oplus$ hot Earth on a 0.84-day orbit, and a much more distant planet of radius 2.4 R$_\oplus$ with an orbital period of 45.3 days (a period ratio well beyond 50:1).  Yet another possibility are systems like 55 Cancri \citep{Fischer:2008,Dawson:2010}, which has an isolated hot Earth in a richer multiplanet system.  Here the period ratio between the innermost planet and its nearest neighbor is nearer to 20---about half the period ratio observed in Kepler-10.

\begin{figure}
\centering
\includegraphics[width=.9\linewidth]{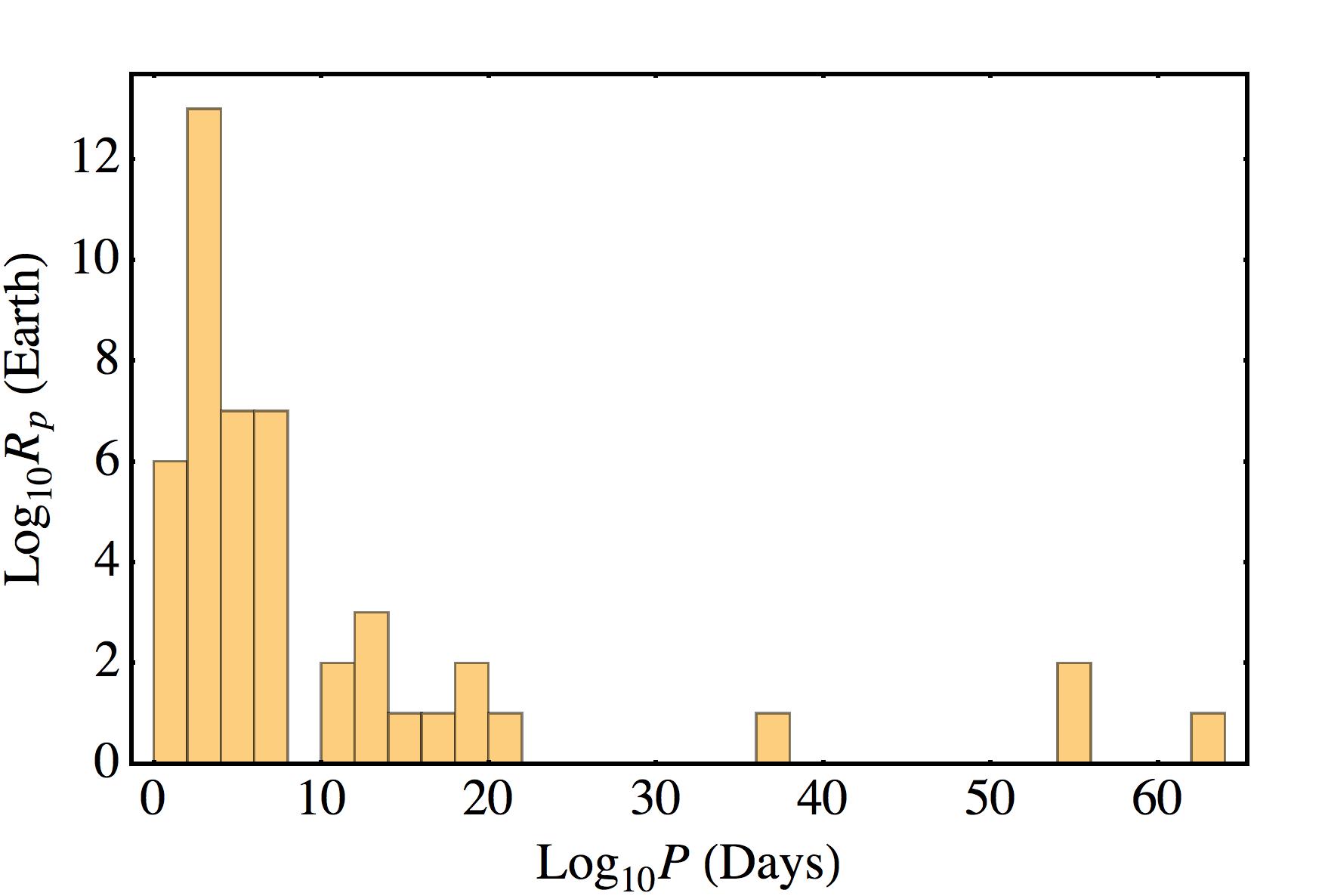}
\caption{Orbital period ratios of observed adjacent planet pairs from multiplanet systems that contain a hot Earth.  For each planet pair with a period ratio of several tens, we estimate there may be $\sim 15$ observed, single hot Earths with a similarly distant companion planet.}
\label{pratios}
\end{figure}

Figure \ref{pratios} shows the observed period ratios for hot Earths in known \kepler\ multiplanet systems.  Four pairs, including Kepler-10, have period ratios larger than 30 and several more have period ratios larger than 10.  Using the same analysis used for the previous section (the number of hot Earth's in standard \kepler\ multis), we estimate that there may be as many as $\sim 15$ hot Earths with undetected outer planets for each system like Kepler-10 (i.e., systems with an inner hot Earth and an observed outer planet with a similarly large ratio of orbital periods).

\section*{Discussion}

If the excess hot Earths are remnant hot Jupiter cores, then it implies that the number of hot Jupiter progenitor systems could be roughly a factor of two larger than what we infer from the current estimates of the frequency of hot Jupiters.  In this scenario the architectures of the outer parts of the systems would match those of the hot Jupiters.  The timescale for hot Earth production via Roche lobe overflow may be a fair fraction of the lifetime of solar-like stars (though different assumptions can yield different results \citep{Valsecchi:2014}).  An Anderson-Darling test comparing the stellar ages used in the \kepler\ false positive analysis for the hot Earth sample with the ages for all planet candidates can not establish an age difference with high confidence.  A more thorough comparison of the host stars of the hot Jupiters and the hosts of hot Earths may yield more insight into the viability of this mechanism, but lies beyond the scope of this work.

Alternatively, a means to produce systems with the architecture of Kepler-10 was suggested by Terquem \citep{Terquem:2014}.  Here, the objects form in a massive disk, beginning at orbital distances of a few AU.  The inner objects start their inward migration before their more distant planetary siblings.  The migration of each planet stops as the inner cavity of the disk sweeps outward.  These simulations were able to reproduce the Kepler-10 architecture---in many cases with additional planets beyond the inner two, and often in or near mean-motion resonance.  Such an architecture yields clear observational signatures for ground-based Doppler observations or for studies of Transit Timing Variations (TTVs)---orbital period variations due to planet-planet gravitational interactions \citep{Agol:2005,Holman:2005}.

Follow-up studies of Kepler-10 using Doppler data from both Keck HIRES and HARPS-N, show no evidence of additional planets between Kepler-10b and Kepler-10c \citep{Dumusque:2014,Weiss:2016}.  At the same time, Weiss et al. \citep{Weiss:2016} identify some evidence for an additional planet in this system from TTVs---specifically TTVs of the outer planet, Kepler-10c.  One possible location for this third planet does lie between the two known planets, at an orbital period near 24 days (still having a very large period ratio with respect to the innermost planet).  However, their preferred model is for the third planet to have an orbit near 100 days---exterior to Kepler-10c and at a period ratio near 2.2 (which itself is interesting due to its frequent appearance in the distribution of period ratios and its relatively large distance from resonance \citep{Baruteau:2013,Steffen:2015}).

Finally, a third mechanism, outlined by Schlaufman et al. \citep{Schlaufman:2010}, predicts an abundance of hot Earths in multiplanet systems where dynamical interactions among the planets drives the innermost planet close to the host star.  These planets may survive for a large fraction of the main sequence lifetime of the host star provided the tides in the host are not highly dissipative (as would be the case for hotter, primarily radiative stars).  In our hot Earth sample, the distribution of stellar effective temperatures differs significantly from the distribution of temperatures of all planet candidates (an Anderson-Darling test gives a $p$-value near $10^{-6}$).  However, the temperatures of the host stars are systematically lower than the planet candidate catalog as a whole.  This observation does not preclude this third mechanism, but it may require a more detailed modeling between the planet and the host star to reconcile this discrepancy.

Many of the planet candidates that we classify as hot Earths also satisfy the criteria for ``Ultra Short Period'' planets studied in \citep{Sanchis-Ojeda:2014}, and many are included in their sample.  Their systems, which are characterized by planets with orbital periods less than one day, adhere to the apparent widening of the period ratios as the inner planets approach the host star \citep{Steffen:2013c}.  They also suggest that most (or all) of their planets likely have outer companions with orbital periods less than $\sim 50$ days.  Both statements are consistent with the sample we study (given systems like Kepler-10).  However, while many of their systems are likely to be members of the population that we claim to identify in this work, it is not clear that their systems will universally be members.  The reason such a claim cannot be made is that the physical mechanism that produces these new systems has not been determined---thus, no comparison can be made with the appropriate theoretical predictions.  Nevertheless, we expect planets in their sample to be members at least as often as planets in ours ($\gtrsim 17\%$).

Characterizing the full architectures of hot Earth systems and determining the true rate of systems with this architecture should give clues regarding their formation and dynamical histories and how those histories differ from standard \kepler\ multiplanet systems and from the Solar System itself.  It may also show that multiple mechanisms may contribute to the population.  Unfortunately, the stars hosting our sample of hot Earths tend to have magnitudes $K_p \gtrsim 14$, with Kepler-10 being the notable exception at $K_p = 11$.  Their dimness renders them difficult to investigate from the ground.  Similarly, among the \nhots\ single hot Earths there are only 10 with $K_p < 13$.  Nevertheless, we expect to find a large number of similar hot Earth systems from the K2 and TESS missions \citep{Howell:2014,Ricker:2014}, whose much brighter targets are more amenable to Doppler follow-up observations and may therefore yield the essential information.

This population of systems need not be restricted to those with a single hot Earth.  Other systems could follow the same dynamical path, and have isolated planets predominantly appearing as single-planet systems, but with different sizes and orbital periods than the sample used to identify the population here.  The existence of this population may contribute to the putative overabundance of single planets among the \kepler\ systems \citep{Moriarty:2015,Hansen:2013,Tremaine:2012,Lissauer:2011b}.  A campaign to study hot Earths discovered by K2 and TESS would yield valuable information about the true architectures of these systems and could provide a robust criteria to determine whether or not a given planetary system followed the associated dynamical path.

\section*{Conclusions}

We have shown that a sizeable portion (at least one in six and perhaps over 40\%, see Table \ref{budgettable}) of Earth-sized planets on one-day orbits among the \kepler\ planet candidates must be from planetary systems with a unique system architecture---and hence have a unique formation or dynamical history.  Of the \nhots\ hot Earths, we estimate that fewer than 10 are due to astrophysical or statistical false positives.  At the same time, we estimate that fewer than \nmult\ of the hot Earths represent the innermost planets of typical multiplanet systems.  Thus, more than \nnew\ of the hot Earths must be members of this new population.  Some of these systems, or systems of similar origin, may contribute to the claimed excess of single-planet systems in the \kepler\ sample \citep{Lissauer:2011b,Hansen:2013,Moriarty:2015}.  (We note corrections to the orbital period of five hot Earth candidates.)

Overall, the rate of these new systems is similar in scale to the rate of hot Jupiters.  In our sample of planet candidates there are 79 hot Jupiters---single planet candidates with radii $r > 6 \ \textrm{R}_\oplus$ and orbital periods $P<7$ days.  Earth-sized planets are more difficult to detect than Jupiters.  We can estimate the detection rate of hot Earths using the results of the TERRA pipeline completeness test---which uses transit signal injection and recovery \citep{Petigura:2013a}.  The semi-analytic approximation to the TERRA completeness from \citep{Steffen:2015} yields the estimate that the hot Earth sample is only $\sim 50\%$ complete, a value that agrees with the \kepler\ Q1-Q17~DR24 injection experiment \citep{Christiansen:2015}. This fact implies that the true number of these systems is roughly twice the observed number ($\gtrsim 50$ systems).

At least three models exist for forming an isolated hot Earth---remnant cores of hot Jupiters \citep{Valsecchi:2014}, systems where inner planets run away from their siblings in a gas disk \citep{Terquem:2014} and are therefore observed as single, and systems where the innermost planet is scattered to an isolated, inner orbit \citep{Schlaufman:2010}.  Still other models may be possible explanations such as some other effect of the protoplanetary disk, or a change in the mutual inclination of orbits at short orbital periods (e.g., due to star-planet interactions).  The divergence in observed orbital period ratios for planets with orbits less than a few days \citep{Steffen:2013c} may also result from the physics driving the formation of these systems.  The initial sizes of the hot Earths may also have been larger---originally being sub-Neptune planets where atmospheric evaporation is responsible for their smaller size compared to most \kepler\ planets \citep{Valencia:2010,Lopez:2013a,Owen:2013}.  New observations regarding the properties of these systems and their host stars will shed light on their origins and give insights into the various branch points in the evolution of planetary systems.  We encourage the exploration of additional models that can be compared to existing and expected observational data.  Anticipated discoveries by the TESS mission \citep{Ricker:2014} will be particularly valuable since the target stars will be bright enough for a robust ground-based, follow-up campaign to establish more completely the architectures of the systems.


\section*{Appendices}

\subsection*{Materials and Methods}

The calculations in this work were done using Mathematica.  The Kernel Density Estimates for the distribution of planet radius and orbital period and the smooth histogram of false positives used a Gaussian kernel with a bandwidth selected using the Silverman method (these are default settings in Mathematica).

To correct for pipeline completeness when we estimate the probability of detecting an outer planet given the observation of a hot Earth, we use the completeness estimates for the TERRA pipeline \citep{Petigura:2013a} (which has better completeness information at short orbital periods than the \kepler\ pipeline \citep{Christiansen:2015}---where the focus was on longer-period planets).  While the absolute sensitivities of the two pipelines differ, the relative sensitivity as a function of orbital period are very similar (lines of constant detection probability have very similar power-law indices).  We generate a simulated planet with a radius 12\% larger than the observed planet.  Then, in order to remove the effects of different sensitivities between the \kepler\ and TERRA pipelines, we calculate the ratio of the detection probability of that simulated planet at a simulated orbital period to the probability of detecting the simulated planet planet at the location of the observed hot Earth.  This approach assumes that the slightly larger planet would have been detected at the location of the hot Earth.


\subsection*{Acknowledgements}
J.H.S. is supported by NASA under grant NNH12ZDA001N-KPS issued through the Kepler Participating Scientist Program and grant NNH13ZDA001N-OSS issued through the Origins of Solar Systems program.  He thanks the \kepler\ multibody working group, Jason Rowe, Jason Hwang, and Chris Burke for ongoing discussions and the members of the Kepler Science Office for their work preparing the Kepler data products and catalogs.  J.L.C. is supported by NASA grant NNX13AD01A.  This research made use of the NASA Exoplanet Archive, which is operated by the California Institute of Technology, under contract with NASA under the Exoplanet Exploration Program.




\bibliographystyle{plainnat}
\bibliography{multis}

\label{lastpage}

\end{document}